\documentclass[11pt]{article}
\usepackage{amsfonts,amsmath,amssymb}
\usepackage{cite}
\usepackage{graphicx} 
\usepackage{mathtools}
\usepackage[margin=0.8in]{geometry}

\usepackage{boondox-cal}
\usepackage[english]{babel}
\usepackage{cite}

\usepackage{xcolor}

\usepackage[utf8]{inputenc}
\usepackage{slashed}

\newcommand{\be}{\begin{equation}}
\newcommand{\bea}{\begin{eqnarray}}
\newcommand{\ba}{\begin{align}}
\newcommand{\ee}{\end{equation}}
\newcommand{\eea}{\end{eqnarray}}
\newcommand{\ea}{\end{align}}

\def\1eq#1{Eq.~(\ref{#1})}

\def\2eqs#1#2{Eqs.~(\ref{#1}) and~(\ref{#2})}
\def\3eqs#1#2#3{Eqs.~(\ref{#1}),~(\ref{#2}) and~(\ref{#3})}
\def\4eqs#1#2#3#4{Eqs.~(\ref{#1}),~(\ref{#2}),~(\ref{#3}) and~(\ref{#4})}

\def\s#1{{\scriptscriptstyle #1}}

\def\G{\Gamma}

\def\s{\mathcal{s}}

\def\tomega{\widetilde{\Omega}}
\def\pomega{{\Omega'}}
\def\pphi#1{{\phi'_#1}}
\def\hAst{{\widehat{A}^*}}
\def\hphist{{\widehat{\phi}^*}}

\def\hphi0{{\hat\phi}_0}

\def\d{\!\mathrm{d}^4x\,}

\def\cf{\epsilon}

\def\user@resume{resume}
\def\user@intermezzo{intermezzo}
\newcounter{previousequation}
\newcounter{lastsubequation}
\newcounter{savedparentequation}
\makeatother

\begin{document}
\
\vskip 3 truecm

\begin{center}
    \bf 
    \Large
    Slavnov-Taylor Identities in Spontaneously Broken\\ Non-Abelian Effective Gauge Theories
\end{center}

\vskip 1 truecm
\begin{center}
    A.~Quadri\footnote{e-mail address: {\tt andrea.quadri@mi.infn.it}}  \\
    INFN, Sezione di Milano\\
    via Celoria 16, I-20133 Milano, Italy
\end{center}



\vskip 1 truecm

\noindent
We study the solution to the Slavnov-Taylor (ST) identities
in spontaneously broken effective gauge theories for a non-Abelian
gauge group. The procedure to extract the $\beta$-functions of 
the theory in the presence of (generalized) non-polynomial field
redefinitions is elucidated.
%

\vskip 10 truecm
{\it 
\small
\noindent
Contributed to the Proceedings of the Steklov Institute of Mathematics.\\
Special volume in honour of Prof.~A.~A.~Slavnov on his 80th anniversary}
\vskip 1 truecm




\vfill\eject
\section{Introduction}

The Slavnov-Taylor (ST) identities ~\cite{Slavnov:1972fg} play a pivotal role
in the consistent quantization of gauge theories, ensuring
the fulfillment of physical unitarity~\cite{Kugo:1977zq,Kugo:1977yx,Becchi:1974xu,Curci:1976yb,Ferrari:2004pd} to all orders in the 
loop expansion.
In the last 50 years ST identities have been thoroughly used in the analysis of
power-counting renormalizable gauge theories~\cite{tHooft:1972tcz,Lee:1973fn}, culminating
in the proof of the all-order renormalizabilty~\cite{Grassi:1995wr,Kraus:1997bi} of the
Standard Model (SM)~\cite{Glashow:1961tr,Weinberg:1967tq,Salam:1964ry} as well as of some of its supersymmetric extensions, most notably the Minimal Supersymmetric Standard Model~\cite{Hollik:2002mv}.

The discovery of the 
Higgs boson at the LHC in 2012~\cite{Chatrchyan:2012xdj,Aad:2012tfa} has confirmed experimentally the SM
by providing evidence of its last missing particle after a search of almost half a century.
On the other hand, up to now no direct evidence has been found at the LHC of supersymmetric particles
at the TeV scale. Moreover no  beyond-the-Standard-Model (BSM) physics effects 
have so far  been detected.

BSM searches are amongst the most challenging ones and new techniques and tools are currently being developed  in order to improve on them (for a recent review see e.g.~\cite{Vagnoni:2019muz}).
It is therefore important to keep an open mind and allow in the analysis of LHC data
for all possible
interactions, consistent with the low energy symmetry pattern of the
spontaneously broken electroweak SU(2)$\times$U(1) group. This can be achieved in the so-called
effective field theory (EFT) approach~\cite{deFlorian:2016spz}
by adding higher dimensional operators arranged in inverse powers of some large energy scale $\Lambda$.
These higher dimensional operators
can be projected on a basis
classified in~\cite{Buchmuller:1985jz,Grzadkowski:2010es}, 
where equations of motion are taken into account in order to
identify on-shell independent
interactions.

At one loop order the full set of anomalous dimensions for the Standard Model EFTs (SMEFTs)  is known~\cite{Jenkins:2013zja,Jenkins:2013wua,Alonso:2013hga}.
Several surprising cancellations in the one-loop UV divergences 
have been observed,  whose origin has been explained in terms of holomorphicity~\cite{Cheung:2015aba,Alonso:2014rga}, and/or remnants of embedding supersymmetry~\cite{Elias-Miro:2014eia}.
It should be noticed that these computations are restricted to be on-shell, since the equations of motion are imposed. This is consistent whenever one is interested in physical S-matrix elements or other physical gauge invariant quantities at one loop order. 

On the other hand, it has been  known since a long time~\cite{Gomis:1995jp} that
such models are indeed renormalizable in the modern sense, i.e. all the UV divergences can be removed order by order in the loop expansion by 
implementing
generalized (usually non polynomial) field redefinitions,
 respecting the ST identities or equivalently the Batalin-Vilkovisky~\cite{Gomis:1994he} master equation, and by redefining the couplings of all possible gauge invariant operators, compatible with the defining symmetries of the theory.

One of the main difficulties in carrying out this task is that 
derivative interactions, which are a characteristic feature of SMEFTs,
maximally violate the power-counting already at one loop order.
For instance  the dimension $6$ operator 
\begin{align}
    \phi^\dagger \phi (D^\mu \phi) D_\mu \phi \, , \quad D_\mu = \partial_\mu -  i A_{a\mu} \frac{\tau_a}{2} \, , \quad
    \phi = \begin{pmatrix} i \phi_1 + \phi_2 \cr \phi_0 - i \phi_3 \end{pmatrix} \, , \quad \phi_0 \equiv v + \sigma \, ,
 \end{align}
$\phi$ being the scalar Higgs doublet, $D_\mu$ the SU(2) covariant derivative with $A_{a\mu}$ the gauge fields and $\tau_a$ the Pauli matrices, and $v$ the vacuum expectation value (v.e.v)
of the $\phi_0$ component, gives rise among others to the interaction vertex $\sim \sigma \partial^\mu \sigma \partial_\mu \sigma$. Hence one can construct a one-loop UV divergent amplitude with an arbitrary number of external $\sigma$-legs, since each interaction vertex contributes two powers of the momenta in the internal loop that compensate the $1/p^2$ behaviour of the scalar propagator.

A constructive strategy for how to obtain the correct field redefintions and carry out the renormalization of the gauge-invariant operators in a scheme-independent way, while respecting the symmetries of the theory, has been recently worked out~\cite{Binosi:2019olm,Binosi:2019nwz,Binosi:2017ubk,Quadri:2016wwl}
within the Algebraic Renormalization approach~\cite{Piguet:1995er,Ferrari:1999nj,Grassi:1999tp,Grassi:2001zz,Quadri:2003ui,Quadri:2003pq,Quadri:2005pv,Bettinelli:2007tq,Bettinelli:2007cy,Bettinelli:2008ey,Bettinelli:2008qn,Anselmi:2012qy,Anselmi:2012jt,Anselmi:2012aq}. 

The problem of maximal power-counting violation can be overcome by using a convenient gauge invariant parameterization of the physical scalar by means of a field $v X_2 \sim  \phi^\dagger \phi - \frac{v^2}{2}$, the constraint being implemented by a suitable Lagrange multiplier field $X_1$. It turns out that the corresponding theory exhibits a (weak) power-counting~\cite{Ferrari:2005va} for the quantized fields (i.e. there is a finite number of divergent amplitudes at each loop order $n$, although this number increases with $n$, as a consequence of the absence of power-counting renormalizability).
We name this auxiliary model the $X$-theory, as opposed to the 
ordinary $\phi$-theory one is eventually interested in.

In the $X$-theory all higher dimensional operators in the classical action are required to vanish at $X_2=0$. Hence the operator $\frac{g}{v\Lambda} \phi^\dagger \phi (D^\mu \phi) D_\mu \phi$  will be replaced by $\frac{g}{\Lambda}X_2 (D^\mu \phi) D_\mu \phi$.
Green's functions of the $\phi$-theory are recovered  by going on-shell with the fields $X_1$ and $X_2$.

Furthermore a suitable set of external sources is introduced in order to formulate in a mathematically consistent way 
the defining functional identities to be fulfilled by the vertex
functional $\G$ of the $X$-theory. In particular it turns out
that the 1-PI Green's functions depending on $X_{1,2}$ are fixed
by amplitudes involving external sources and insertions of
quantized fields other than $X_{1,2}$~\cite{Binosi:2019olm}.
We name the latter amplitudes ancestor Green's functions.

Once the renormalization of the $X$-theory is achieved, one goes on-shell with the $X_{1}$ and $X_2$, which amounts to apply a suitable mapping of the external sources  onto operators depending on $\phi$ and its covariant derivatives~\cite{Binosi:2017ubk,Binosi:2019olm}. 
This procedure yields the full set of UV divergences of the SMEFT.
Since we are working off-shell 
with respect to (w.r.t.) the field $\phi$ and  other quantized fields in the $\phi$-theory, generalized field redefinitions, that are present already at one-loop order and are in general not even polynomial also in the Abelian case~\cite{Binosi:2017ubk,Binosi:2019nwz},
are automatically accounted for
through the so-called cohomologically trivial invariants.

The paper is organized as follows.  Notations and conventions are presented in Sect.~\ref{sec.not}.
The solution to the ST identities is described in Sect.~\ref{sec:sol.sti} for a non-Abelian SU(2)
gauge group.
We present both the non-local solution to the ST identity for the full vertex functional $\G$ as well as
 its local approximation (relevant for the recursive parameterization of the UV divergences order by order in the loop expansion).
We then discuss the running of the
couplings by presenting the Renormalizaton Group equation. We finally elucidate how to obtain the $\beta$-functions in the present formalism.

\section{Notations and setup}\label{sec.not}

We consider the $\mbox{SU(2)}$ gauge group. 
The field strength is defined by ($A_\mu = A^a_\mu \frac{\tau_a}{2}$)
\begin{align}
    G_{\mu\nu}[A]= G_{a\mu\nu} \frac{\tau_a}{2} = 
    \partial_{\mu} A_\nu - 
    \partial_{\nu} A_\mu - i
    [ A_\mu, A_\nu ] \, .
\end{align}
It is convenient to introduce a matrix
representation for the scalar $\phi$
by setting
\begin{align}
    \Omega \equiv \frac{1}{2} \Big (
    \phi_0 + i \tau_a \phi_a \Big )  \, .
\end{align}
Notice that $\Omega^\dagger \Omega =
\frac{1}{4}
 ( \phi_0^2 + \phi_a^2) {\bf I}$, where
 ${\bf I}$ is the identity matrix.
Moreover $\phi_0 = v + \sigma$,
$v$ being the vacuum expectation value.

The classical action of the theory is obtained by extending the construction in~\cite{Binosi:2017ubk} to the non-Abelian gauge group. For that purpose one sets
\begin{align}
    S = \int \d & \Big \{  
    -\frac{1}{4g^2} G_{a\mu\nu} G_a^{\mu\nu} +  {\rm Tr} (D_\mu \Omega)^\dagger D^\mu \Omega -
    \frac{M^2- m^2}{2} X_2^2 
    - \frac{m^2}{2v^2} 
    \Big ( {\rm Tr}~ \Omega^\dagger \Omega - 
    \frac{v^2}{2}  \Big )^2 \nonumber \\
    & - \bar c (\square + m^2) c
    + \frac{1}{v} (X_1 + X_2)
    (\square + m^2) \Big ( 
    {\rm Tr}~ \Omega^\dagger \Omega - 
    \frac{v^2}{2} - v X_2
    \Big ) \nonumber \\
    & + \frac{g}{\Lambda} X_2
    {\rm Tr} (D_\mu \Omega)^\dagger D^\mu \Omega + T_1 {\rm Tr} (D_\mu \Omega)^\dagger D^\mu \Omega
    \Big \} \, ,
    \label{cl.act}
\end{align}
$g$ being the SU(2) coupling constant.
By going on-shell with the field
$X_1$ (that plays the role of a Lagrange multiplier)
one recovers the constraint\footnote{
Going on-shell with $X_1$
yields 
the condition
\begin{align}
    (\square + m^2) \Big (
      {\rm Tr}~ \Omega^\dagger \Omega - \frac{v^2}{2} - v X_2 \Big ) = 0 \, ,
\end{align}
whose most general solution is
$X_2 = \frac{1}{v} \Big (
    {\rm Tr}~ \Omega^\dagger \Omega - \frac{v^2}{2} \Big ) + \eta,$ $\eta$ being a
    scalar field of mass $m$.
In perturbation theory
the correlators of the mode $\eta$ with any gauge-invariant operators vanish~\cite{Binosi:2019olm}, hence one can safely set $\eta =0$.
} 
\begin{align}
    v X_2 = {\rm Tr}~ \Omega^\dagger \Omega - 
    \frac{v^2}{2} \, ,
\end{align}
which, when substituted back into Eq.(\ref{cl.act}), yields the usual quartic potential with coupling~$\sim M^2$, namely
$$-\frac{M^2}{2}
\Big ( {\rm Tr}~ \Omega^\dagger \Omega - \frac{v^2}{2} \Big )^2 \, .$$
The physical scalar excitation 
has mass $M$. 
We observe that the classical action 
in Eq.(\ref{cl.act}) depends on a parameter~$m^2$ associated with the quartic potential of the field $\phi$ that compensates with the contribution from the quadratic mass term for $X_2$ once one goes on-shell with $X_{1,2}$. It follows that Green's functions in the target theory have to be $m^2$-independent. This condition 
turns out to be a very strong check of the computations, since the parameter~$m^2$ enters
non-trivially both in the invariants and the Feynman amplitudes.

The last line of Eq.(\ref{cl.act}) contains the dim.6 derivative operator and the external source $T_1$, required to define the 
$X_2$-equation at the quantum level, as we will soon discuss.

The classical action $(\ref{cl.act})$
is invariant under two distinct BRST symmetries. The first is the U(1) {\it constraint} BRST symmetry, acting as follows:
\begin{align}
    \s X_1 = v c \, , \qquad \s c = 0 \, , \qquad \s \bar c = {\rm Tr}~ \Omega^\dagger \Omega - 
    \frac{v^2}{2} - v X_2 \, ,
\end{align}
while leaving all other fields and external sources invariant.
It ensures that no additional physical degree of freedom is introduced in the $X$-formalism~\cite{Quadri:2006hr,Binosi:2017ubk,Binosi:2019olm,Quadri:2016wwl}. 
The U(1) constraint ghost and antighost fields remain free.

In addition the classical action is invariant under the SU(2) BRST symmetry, acting as follows (we denote the SU(2) ghosts by $\omega_a$):
\begin{align}
& s A_{a\mu} = \partial_\mu \omega_a + \cf_{abc} A_{b\mu} \omega_c \, , 
\qquad  s \omega_a = -\frac{1}{2} \cf_{abc} \omega_b \omega_c \, , \nonumber \\
& s \phi_0 = -\frac{1}{2} \omega_a \phi_a\,,  \qquad  s \phi_a = \frac{1}{2} \phi_0 \omega_a  + \frac{1}{2} \cf_{abc} \phi_b \omega_c \, ,
\label{su2.brst}
\end{align}
with $\cf_{abc}$ the SU(2) structure constant.
$\s$ and $s$ anticommute and are both nilpotent.

Gauge-fixing is carried out in the usual way {\em \`a la} BRST by introducing the antighosts $\bar \omega_a$ paired into a BRST doublet~\cite{Quadri:2002nh,Barnich:2000zw} with the Nakanishi-Lautrup fields $b_a$:
\begin{align}
    s \bar \omega_a = b_a \, , \qquad s b_a = 0 \, .
\end{align}
We adopt a standard $R_\xi$-gauge and introduce the gauge-fixing and ghost action as
\begin{align}
    S_{\rm g.f} + S_{\rm ghost} =
    \int \d \Big [ 
    \frac{b_a^2}{2\xi} - b_a \Big (
    \partial A_a + \frac{v}{\xi} \phi_a
    \Big )
    + \bar \omega_a \partial^\mu (D_\mu[A] \omega)_a + \frac{v}{2 \xi} \bar \omega_a (\phi_0 \omega_a + \cf_{abc} \phi_b \omega_c )
    \Big ] \, .
\end{align}
$(D_\mu[A]\omega)_a = \partial_\mu \omega_a + \cf_{abc} A_{b\mu} \omega_c$
is the SU(2) covariant derivative.
We also need to introduce
the antifields associated with the $s, \s$-BRST
transformations that are non-linear in the quantized fields:
\begin{align}
    S_{\rm a.f.} = \int \d \Big [
    A^*_{a\mu} (D_\mu[A] \omega)_a - 
    \frac{1}{2} \sigma^* \omega_a \phi_a + \frac{1}{2} \phi_a^* 
    (\phi_0 \omega_a + \cf_{abc} \phi_b \omega_c ) + \bar c^*
    \Big ( {\rm Tr} ~\Omega^\dagger \Omega - \frac{v^2}{2} - v X_2\Big )
    \Big ] \, .
\end{align}
The complete tree-level vertex functional is given by
\begin{align}
    \G^{(0)} &= S +  S_{\rm g.f} + S_{\rm ghost} +  S_{\rm a.f.} \, .
\end{align}
It obeys the following set of functional identities:
\begin{enumerate}
\item The $b$-equation
\begin{align}
    \frac{\delta \G^{(0)}}{\delta b_a}=\frac{b_a}{\xi} - \partial A_a - \frac{v}{\xi} \phi_a \, ;
    \label{b.eq}
\end{align}
\item The SU(2) antighost equation
\begin{align}
    \frac{\delta \G^{(0)}}{\delta \bar \omega_a} = \partial^\mu \frac{\delta \G^{(0)}}{\delta A^*_{a\mu}} + \frac{v}{\xi}
    \frac{\delta \G^{(0)}}{\delta \phi_a^*} ;
    \label{su2.ag.eq}
    \end{align}
\item The constraint U(1) ghost and antighost equations
\begin{align}
   \frac{\delta \G^{(0)}}{\delta  c} = (\square + m^2) \bar c \, ; 
   \qquad
   \frac{\delta \G^{(0)}}{\delta \bar c} = - (\square + m^2) c \, ; 
   \label{u1.gh.ag.eq}
\end{align}
\item The $X_1$-equation
\begin{align}
 \frac{\delta \G^{(0)}}{\delta X_1} = 
 \frac{1}{v} (\square + m^2)
 \frac{\delta \G^{(0)}}{\delta \bar c^*} ;
 \label{x1.eq}
\end{align}
\item The $X_2$-equation
\begin{align}
 \frac{\delta \G^{(0)}}{\delta X_2} =
 \frac{1}{v} (\square + m^2)
 \frac{\delta \G^{(0)}}{\delta \bar c^*} + \frac{g}{\Lambda}
 \frac{\delta \G^{(0)}}{\delta T_1}
 - (\square + M^2) X_2 - (\square + m^2) X_1 - v \bar c^* \, ;
 \label{x2.eq}
\end{align}
\item The ST identity
\begin{align}
    {\cal S}(\G^{(0)}) \equiv
    \int \d \Big [ \frac{\delta \G^{(0)}}{\delta A^*_{a\mu}} 
    \frac{\delta \G^{(0)}}{\delta A_{a\mu}} + 
    \frac{\delta \G^{(0)}}{\delta \sigma^*} 
    \frac{\delta \G^{(0)}}{\delta \sigma} + 
    \frac{\delta \G^{(0)}}{\delta \phi_a^*} 
    \frac{\delta \G^{(0)}}{\delta \phi_a} +
    \frac{\delta \G^{(0)}}{\delta \omega_a^*} 
    \frac{\delta \G^{(0)}}{\delta \omega_a} 
    + b_a     \frac{\delta \G^{(0)}}{\delta \bar \omega_a} 
    \Big ] = 0 \, .
    \label{st.eq}
\end{align}
\end{enumerate}
Since the gauge group is non-anomalous,
these functional identities can be preserved to all orders in the loop expansion.
We remark that the ST identity for 
the constraint U(1) BRST symmetry
\begin{align}
    {\cal S}_{\scriptscriptstyle{C}}(\G^{(0)}) \equiv \int \!\mathrm{d}^4 x \, \Big [ v c \frac{\delta \G^{(0)}}{\delta X_1} 
 + \frac{\delta \G^{(0)} }{\delta \bar c^*}\frac{\delta \G^{(0)} }{\delta \bar c} \Big ] = 
 \int \!\mathrm{d}^4 x \, \Big [ v c \frac{\delta \G^{(0)}}{\delta X_1} 
 -(\square + m^2) c \frac{\delta \G^{(0)}}{\delta \bar c^*} \Big ] = 0
 \label{sti.c} 
\end{align}
is not an independent equation but reduces to the $X_1$-equation of motion (\ref{x1.eq}), the ghost $c$ being a free field.

\section{Solutions of the ST identity}
\label{sec:sol.sti}

We expand the vertex functional $\G$ according to the loop order.
$\G^{(n)}$ denotes the $n$-th order
coefficient in such an expansion (the generating functional of $n$-th loop 1-PI amplitudes).

\subsection{Change of variables}

In order to solve the set of functional
equations (\ref{b.eq})-(\ref{st.eq}) 
for $\G^{(n)}$ we introduce a suitable set of external sources redefinitions that automatically takes into account all functional identities but the ST identity.
The $b$-equation (\ref{b.eq}) at order $n \geq 1$ in the loop expansion reads
\begin{align}
    \frac{\delta \G^{(n)}}{\delta b_a} = 0 \, ,
    \label{b.n}
\end{align}
stating that the whole dependence on the Nakanishi-Lautrup field $b_a$ is confined at tree-level.
Similarly the constraint U(1) ghost and antighost equations (\ref{u1.gh.ag.eq}) entail that $\G^{(n)}$, $n \geq 1$ does not depend on $\bar c, c$.
By the SU(2) antighost equation the dependence on the antighost field $\bar \omega_a$ only appears in the combinations
\begin{align}
\widehat A^*_{a\mu} = A^*_{a\mu} - \partial_\mu \bar \omega_a \, , \qquad
\widehat \phi^*_a = \phi^*_a + \frac{v}{\xi} \bar \omega_a \, .
\label{af.redef}
\end{align}
The $X_{1,2}$-equations are finally solved
by the replacement
\begin{align}
    {\cal T}_1 = T_1 + \frac{g}{\Lambda} X_2 \, , 
    \qquad
    \bar {\cal c}^* = 
    {\bar c}^* + \frac{1}{v}
    (\square + m^2) (X_1 + X_2) \, ,
\end{align}
as it can be seen by applying the chain rule for functional differentiation.


\subsection{The $n$-th order ST identity}

We now need to solve the ST identity. The procedure is a recursive one, order by order in the loop expansion.
The $n$-th order ST identity reads
\begin{align}
    {\cal S}_0(\G^{(n)}) + \Delta^{(n)} =0,
    \label{nloop.st}
\end{align}
where $\Delta^{(n)}$ takes into account the lower order contributions
\begin{align}
    \Delta^{(n)} = \sum_{j=1}^{n-1}    \int \d  \Big [
     \frac{\delta \G^{(n-j)}}{\delta \hAst_{a\mu}} 
    \frac{\delta \G^{(j)}}{\delta A_{a\mu}} + 
    \frac{\delta \G^{(n-j)}}{\delta \sigma^*} 
    \frac{\delta \G^{(j)}}{\delta \sigma} + 
    \frac{\delta \G^{(n-j)}}{\delta \hphist_a} 
    \frac{\delta \G^{(j)}}{\delta \phi_a} +
    \frac{\delta \G^{(n-j)}}{\delta \omega_a^*} 
    \frac{\delta \G^{(j)}}{\delta \omega_a} \Big ] \, 
\end{align}
and
${\cal S}_0$ is the linearized ST operator
\begin{align}
	{\cal S}_0  (\G^{(n)}) & = \int \!\mathrm{d}^4 x \, \Big [ (D_\mu[A] \omega)_a \frac{\delta \G^{(n)}}{\delta A_\mu}  + 
	+ \frac{1}{2} ( \phi_0 \omega_a + 
	\cf_{abc} \phi_b \omega_c ) \frac{\delta \G^{(n)}}{\delta \phi_a}  
	-\frac{1}{2} \omega_a \phi_a \frac{\delta \G^{(n)}}{\delta \sigma}
\nonumber \\
&  
+ \frac{\delta \G^{(0)}}{\delta A_{a\mu}} \frac{\delta \G^{(n)}}{\delta \hAst_{a\mu}} 
+ \frac{\delta \G^{(0)}}{\delta \sigma} \frac{\delta \G^{(n)}}{\delta \sigma^*} + \frac{\delta \G^{(0)}}{\delta \phi_a} \frac{\delta \G^{(n)}}{\delta \hphist_a} \Big ] \nonumber \\
& = s \G^{(n)} + \int \!\mathrm{d}^4 x \, \Big [
\frac{\delta \G^{(0)}}{\delta A_{a\mu}}
 \frac{\delta \G^{(n)}}{\delta \hAst_{a\mu}} 
+ \frac{\delta \G^{(0)}}{\delta \sigma} \frac{\delta \G^{(n)}}{\delta \sigma^*} + \frac{\delta \G^{(0)}}{\delta \phi_a} \frac{\delta \G^{(n)}}{\delta \hphist_a}   \Big ],
\label{S0}
\end{align}
which acts as the BRST differential $s$ on the fields of the theory while mapping the antifields into the classical equations of motion of their corresponding fields. Notice that there is no $(\bar \omega_a,b_a)$-dependent term in Eq.(\ref{S0}), since
by Eq.(\ref{b.n}) $\G^{(n)}$ is $b$-independent and
we are using the combinations in Eq.(\ref{af.redef}).

Our goal is to recusrively solve Eq.(\ref{nloop.st}) for the full vertex functional $\G^{(n)}$. So we assume that the ST identity holds true up to order $n-1$ and that $\G^{(j)}$, $j < n$ are known.

The first step is to trivialize the BRST symmetry as much as possible by performing an invertible change of variables leading to either ${\cal S}_0$-invariant combinations or to doublets, i.e. pairs $(u,w)$
such that
${\cal S}_0 u = w\, , {\cal S}_0 w = 0$.

For that purpose let us introduce the SU(2) matrix 
\begin{align}
    \pomega = \pphi{0} + i \pphi{a} \tau_a \equiv \sqrt{\frac{2}{ {\rm Tr} ~ \Omega^\dagger \Omega}} ~ \Omega \, ,  \quad 
    {\pomega}^\dagger \pomega = {\bf 1} \, , \quad
    \pphi{0} = \frac{\phi_0}{\sqrt{\phi_0^2+\phi_a^2}} \, , \quad 
    \pphi{a} = \frac{\phi_a}{\sqrt{\phi_0^2 + \phi_a^2}} \, ,
    \label{bleach}
\end{align}
and then carry out an operatorial gauge-transformation inspired by the
bleaching procedure in the St\"uckelberg model~\cite{Bettinelli:2007tq}:
\begin{align}
    a_\mu & = -i a_{0\mu} {\bf 1} + a_{a \mu} \frac{\tau_a}{2} \equiv
    \pomega^\dagger A_\mu \pomega - i \partial_\mu \pomega^\dagger \pomega \, , \nonumber \\
    \tomega & = \pomega^\dagger \Omega \, .
    \label{bleaching}
\end{align}
Each of the components of
$a_\mu$ and $\tomega$ are both ${\cal S}_0$- (and gauge-) invariant.
Moreover the transformation in Eq.(\ref{bleaching}) is invertible. In components we
find
\begin{align}
a_{a\mu} &=
    (\pphi{0}^2 - \pphi{a}^2) A_{a\mu}
    + 2 \pphi{a} \pphi{b} A_{b\mu} +
    2 \epsilon_{abd} A_{b\mu} \pphi{a} \pphi{0}
    \nonumber \\
    & + 2 (\pphi{a} \partial_\mu \pphi{0} -
    \pphi{0} \partial_\mu \pphi{a}) +
    2 \epsilon_{abc} \partial_\mu \pphi{b} \pphi{c} \, ,
\end{align}
while the only non-vanishing component of $\tomega$ is in one-to-one correspondence with $\sigma$:
\begin{align}
  \tilde{\phi}_0 \equiv {\rm Tr}~ \tomega = 
  \sqrt{\phi_0^2 + \phi_a^2} = 
  v + \sigma + \frac{1}{2} \frac{\phi_a^2}{v^2} ( v- \sigma) + \dots
\end{align}
where the dots stand for terms
of order $\sigma^2$ and $\phi_a^ 2$.
Notice in particular that
\begin{align}
    \left . a_{a\mu} \right |_{\phi_a=0} = A_{a\mu} \, , \qquad
    \left .   \tilde \phi_0 \right |_{\phi_a=0} = v + \sigma \, .
\end{align}
The trace of $a_\mu$ yields
\begin{align}
    a_{0\mu} = \partial_\mu ( \phi_0^2 + \phi_a^2) \, ,
\end{align}
which is trivially invariant.
As a side remark we notice that $a_{0\mu}$ would vanish in the
St\"uckelberg model where the constraint
$\phi_0^2+\phi_a^2=v^2$ holds true.

Moreover we can also redefine the ghost fields
\begin{align}
    \tilde \omega_a \equiv s \phi_a =
    \frac{1}{2} \phi_0 \omega_a + \frac{1}{2} \cf_{abc} \phi_b \omega_c
\end{align}
and again such a transformation is invertible. We stress 
that invertibility is related to the properties of the linearized BRST transformation
of Goldstone fields, that has to be non-vanishing since the Goldsone are unphysical modes cancelling in the quartet mechanism against the ghosts and the unphysical longitudinal
polarizations of the gauge fields~\cite{Becchi:1985bd}.
This is a crucial and general property of spontaneously broken gauge theories.

Under the BRST differential $s$
the new variables are either invariant, namely 
$a_{a\mu}$ and $\tilde \phi_0$,
or coupled into a BRST pair, as
$(\phi_a,\tilde \omega_a)$.

We now move to the antifield sector.
In the $X$-formalism we can redefine
the $\bar c^*$ so that it forms a ${\cal S}_0$-doublet
with $\sigma^*$. This is because
\begin{align}
    {\cal S}_0 (\sigma^*) = 
    \frac{\delta \G^{(0)}}{\delta \sigma} = v \bar c^* + \dots \, ,
\end{align}
so that the redefinition $\bar c^* \rightarrow \widetilde{\bar c}^* \equiv \frac{1}{v} \frac{\delta \G^{(0)}}{\delta \sigma}$
is invertible. 

In a similar way the antifield $\omega_a^*$ pairs 
into a ${\cal S}_0$-doublet:
\begin{align}
{\cal S}_0 ( \omega_a^*) =
\frac{\delta \G^{(0)}}{\delta \omega_a}
= (D^\mu[A] \hat{A^*_\mu})_a + \epsilon_{abc}
\omega^*_b M_{cd} \tilde{\omega}_d
- \frac{1}{2} \hphist_a \phi_0
-\frac{1}{2} \epsilon_{abd} \hphist_b \phi_d
\equiv \widetilde{\phi^*_a} \, .
\end{align}
Again, since $\phi_0 = v + \sigma$,
this transformation is invertible.
In the above equation we have used the matrix
$M_{cd}$ such that
\begin{align}
    & \tilde \omega_a = R_{ac} \omega_c \equiv\Big ( \frac{1}{2} \phi_0 \delta_{ac} + \frac{1}{2} \epsilon_{abc} \phi_b \Big ) \omega_c \, , 
    \quad \nonumber \\
    & \omega_i = M_{ia} \tilde\omega_a \equiv 2 \Big [
     \frac{\phi_0}{\phi_0^2+\phi_a^2}\delta_{ia} +  \frac{\phi_i \phi_a}{\phi_0(\phi_0^2 + \phi_a^2)} +
    \frac{\epsilon_{iaq} \phi_q}{\phi_0^2 + \phi_a^2}
    \Big ] \, , \qquad
    M_{ia} R_{ac} = \delta_{ic} \, .
\end{align}
In the new variables Eq.(\ref{nloop.st}) can  be cast in the form
\begin{align}
    \rho(\G^{(n)}) & \equiv
    \int \d \Big [
    \tilde \omega_a 
    \frac{\delta \G^{(n)}}{\delta \phi_a} + 
    \widetilde{\bar c}^*
    \frac{\delta \G^{(n)}}{\delta \sigma^*}
    +
    \widetilde{\phi^*_a}
    \frac{\delta \G^{(n)}}{\delta \omega^*_a}
    \Big ] \nonumber \\
    & = R^{(n)} \equiv
    - \int \d \Big [ \frac{\delta \G^{(0)}}{\delta A_{a\mu}} \frac{\delta \G^{(n)}}{\delta \hAst_{a\mu}} + \frac{\delta \G^{(0)}}{\delta \phi_a} \frac{\delta \G^{(n)}}{\delta \hphist_a}
    \Big ] - \Delta^{(n)} \,
    \label{rho.sti}
\end{align}
with the help of an auxiliary BRST differential $\rho$ acting only on doublets.
$\rho$ is nilpotent.
It is understood that the last line of the above equation is expressed in terms of
the variables $a^a_\mu, \tilde \phi_0, \tilde \omega_a, \hAst_{a\mu}, \sigma^*, \widetilde{\bar c}^*, \omega_a^*, \widetilde{\phi^*_a} $, $\phi_a, T_1$.

At this point it is convenient to introduce a collective notation for the $\rho$-doublets,
i.e. $u_I = \{ \phi_a, \sigma^*, \omega_a^* \}$, $w_I = \{ \tilde \omega_a, \widetilde{\bar c}^*,\widetilde{\phi^*_a} \}$ so that $\rho u_I = w_I, \rho w_I = 0$.

The important point is that $\rho$ can be inverted since it admits a homotopy operator
${\cal h}$~\cite{Wess:1971yu,Picariello:2001ri}, {\it i.e.}, an operator such that
$\{ \rho, {\cal h} \} = \mathbb{I},$
where $\mathbb{I}$ is the identity on the space of functionals depending on the doublets. Indeed, let us define
\begin{align}
    {\cal h}(Y) = \int_0^1 \!\mathrm{d}t \! \int \d \, 
    \sum_I u_I \lambda_t \frac{\delta Y}{\delta w_I}, 
\end{align}
where the operator $\lambda_t$ acts on a functional $Y$ by multiplying the doublets it depends upon by $t$, without affecting all the other variables $\xi$ on which $Y$ might depend:
\begin{align}
    \lambda_t Y(u_I,v_I;\xi) = 
    Y(t u_I, t v_I ;\xi).
\end{align}
Then one easily verifies that 
\begin{align}
    \{ \rho, {\cal h} \}Y = Y(u_I,w_I;\xi) - Y(0,0;\xi).
\end{align}
This identity holds without locality restrictions,
the only condition being that the space of $Y$ functionals is star-shaped, as
it happens for the space of functionals depending on the fields and the external sources of the theory.

Moreover, as a consequence of the nilpotency of $\rho$, the right hand side (r.h.s.) of Eq.(\ref{rho.sti})
is $\rho$-invariant, and thus one finds
\begin{align}
\rho(\G^{(n)}) = R^{(n)} = \{ \rho,  {\cal h} \} R^{(n)} = \rho {\cal h} R^{(n)}
\label{rho.nil}
\end{align}
since $R^{(n)}$ is $\rho$-invariant.
Eq.(\ref{rho.nil}) yields
the final representation for the $n$-th order vertex functional
\begin{align}
    \G^{(n)} = {\cal h} R^{(n)} + \G^{(n)}_{\mathrm{ker}},
\label{nth.order.rep}
\end{align}
where $\G^{(n)}_{\mathrm{ker}}$ is a $\rho$-invariant functional built from the bleached variables and the $(u_I,w_I)$-doublets.
In particular 
\begin{align}
    \G^{(n)}_{\rm g.i.} = \left . \G^{(n)}_{\mathrm{ker}} \right |_{\tilde \omega = 0} \, .
    \label{nth.order.gi}
\end{align}
is a gauge-invariant functional.

Several comments are in order.
First we remark that Eqs.(\ref{nth.order.rep})  and (\ref{nth.order.gi}) hold for the
complete 1-PI Green's functions, without locality approximations.
In particular the bleaching change of variables (\ref{bleach}) 
dresses the gauge and scalar fields with the appropriate Goldstone dependence required to ensure
gauge invariance. A further dependence on the Goldstone fields is induced by the 
ghost sector (via the homotopy operator) and the variables $\tilde \omega_a,\widetilde{\bar c}^*,\widetilde{\phi^*_a}$.
Notice that at one loop $\Delta^{(n)}$ in Eq.(\ref{rho.sti}) vanishes.

\subsection{Classification of UV divergences}

UV divergences in EFTs are local (in the sense of formal power series)~\cite{Gomis:1995jp}.
In order to classify them one can apply the standard approach based on cohomological tools~\cite{Barnich:2000zw,Gomis:1994he}.
Subtraction of UV divergences is carried recursively in the number of loops.
Suppose that they have been subtracted in a symmetric way, i.e.
fulfilling all the relevant functional identities, up to order $n-1$. 
Then $\Delta^{(n)}$ in Eq.(\ref{nloop.st}) is finite and therefore
the UV-divergent part of $\G^{(n)}$, denoted by $\overline{\G}^{(n)}$,
is ${\cal S}_0$-invariant:
\begin{align}
    {\cal S}_0 (\overline{\G}^{(n)}) = 0 \, .
\end{align}
Due to the nilpotency of ${\cal S}_0$ there are two types of solutions
in the space of local functionals:
gauge-invariant polynomials in the field strength, $\Omega$ and their covariant derivatives (the cohomologically non-trivial sector) and
${\cal S}_0$-exact functionals of the form ${\cal S}_0(Y^{(n)})$
(the cohomologically trivial sector).

The latter are associated with (generalized) field redefinitions.
For instance the redefinition $\sigma \rightarrow \sigma + P^{(n)}(\Phi,\zeta, \partial) \sigma$, where $P^{(n)}$ is a formal power series in the fields
(collectively denoted by $\Phi$) and the external sources (again collectively denoted by $\zeta$) and polynomial in the derivatives, is generated by the invariant
\begin{align}
    \int \d {\cal S}_0 ( P^{(n)}(\Phi,\zeta, \partial) \sigma^*) = 
    \int \d \, \Big [ s P^{(n)} \sigma^* +  P^{(n)}(\Phi,\zeta, \partial) \frac{\delta \G^{(0)}}{\delta \sigma} \Big ] \, .
\end{align}
We emphasize that these invariants connect the ghost-antifield sector (via the first term in the r.h.s. of the above equation) with the contributions induced by the field redefinition on the classical action (the second term).
Indeed one can identify the  field redefinitions actually present by studying the antifield amplitudes.
Explicit computations (in the Abelian case) have shown that such
field redefitions do actually arise already at one loop order and are not even polynomial in the fields~\cite{Binosi:2019olm,Binosi:2019nwz}.

We can parameterize the UV divergences of the theory by the coefficients
$g_i$ of the gauge-invariants~\footnote{In the notation of~\cite{Binosi:2019nwz} $g_i$ include the $\lambda_i$'s  (the coefficients of the gauge invariants only depending on the fields), the $\vartheta_i$'s (coefficients of the invariants only dependent on the external sources) and the $\theta_i$'s (the coefficients of the mixed field-external sources invariants).}, belonging to the cohomologically non-trivial sector, and by the coefficients $\rho_i$ of the cohomologically trivial invariants.

Since a change in the renormalization mass scale $\mu$ can be reabsorbed by a shift in the $g_i$'s and the $\rho_i$'s, we obtain the 
Renormalization Group equation for the model at hand
\begin{align}
    \mu \frac{\partial \G}{\partial \mu} + \mu \frac{\partial g_i}{\partial \mu} \frac{\partial \G}{\partial g_i}  
    + \mu \frac{\partial \rho_i}{\partial \mu} \frac{\partial \G}{\partial \rho_i} = 0 \, .
    \label{RG}
\end{align}
Once the separation between the cohomologically trivial and non-trivial sectors has been correctly implemented by fixing the $\lambda_i$'s and 
the $\rho_i$'s, the derivative w.r.t. $\mu$ of the latter yields the (generalized) anomalous dimensions $\gamma_i$'s of the fields, while the $\beta$-functions are obtained by differentiation of the $g_i$'s:
\begin{align}
\gamma_i \equiv \mu \frac{\partial \rho_i}{\partial \mu} \ , \qquad
\beta_i \equiv \mu \frac{\partial g_i}{\partial \mu} \, .
\end{align}
Notice that when projecting on the Feynman amplitudes, the cohomologically trivial invariants do in general give rise to non-trivial contributions to the UV divergences (due to the non-linearity of the field redefinitions), that have to be taken into account in order to
extract the $g_i$'s and the $\rho_i$'s.
Several explicit examples can be found in~\cite{Binosi:2019nwz}.

\section{Conclusions}\label{sec:concl}

Quantization of spontaneously broken gauge effective
field theories can be consistently carried order
by order in the loop expansion in a full off-shell formulation. The dimensions of the gauge invariants, parameterizing the UV divergences of the $n$-th order vertex functional, then fix the order of
the perturbative expansion in inverse powers 
of the large energy scale $\Lambda$ at order $n$ in the loop expansion.

The consistent recursive subtraction of 
UV divergences of off-shell 1-PI Green's functions is crucial in order to ensure
that overlapping divergences can indeed be removed
by local counter-terms at higher orders in
the perturbative series. Moreover, the fulfillment
of the ST identity has to be guaranteed in order
to respect physical unitarity.

We have shown that in these models the solution
to the ST identity can be written for the full vertex functional $\G$, without local approximations,
by homotopy techniques.
The underlying algebraic structure  is richer than in the power-counting renormalizable case. 

In particular (generalized) field redefinitions have to be taken into account already at one loop order, or else it will not be possible to fix unambiguously the
coefficients $g_i$ of the gauge invariants, separating the genuine physical renormalizations from spurious effects due to the redefinition of the fields.
Such field redefinitions are determined by the anti-field dependent amplitudes.

The explicit renormalization of all dimension 6 operators in non-Abelian gauge theories in the 
formalism described in the present paper is currently under investigation.

\appendix

\section*{Acknowledgments}

It is a pleasure to thank
Prof.~A.A.Slavnov for many enlightening discussions, insights and perspective on a wide range of subjects in renormalization and quantization of gauge theories, as well as for several exciting joint collaborations.

\bibliography{bibliography_1loop.bib}

\begin{thebibliography}{10}

\bibitem{Slavnov:1972fg}
A.~A. Slavnov, ``{Ward Identities in Gauge Theories},'' {\em Theor. Math.
  Phys.}, vol.~10, pp.~99--107, 1972.
\newblock [Teor. Mat. Fiz.10,153(1972)].

\bibitem{Kugo:1977zq}
T.~Kugo and I.~Ojima, ``{Manifestly Covariant Canonical Formulation of
  Yang-Mills Field Theories: Physical State Subsidiary Conditions and Physical
  S Matrix Unitarity},'' {\em Phys. Lett.}, vol.~B73, pp.~459--462, 1978.

\bibitem{Kugo:1977yx}
T.~Kugo and I.~Ojima, ``{Manifestly Covariant Canonical Formulation of
  Yang-Mills Field Theories. 1. The Case of Yang-Mills Fields of Higgs-Kibble
  Type in Landau Gauge},'' {\em Prog. Theor. Phys.}, vol.~60, p.~1869, 1978.

\bibitem{Becchi:1974xu}
C.~Becchi, A.~Rouet, and R.~Stora, ``{The Abelian Higgs-Kibble Model. Unitarity
  of the S Operator},'' {\em Phys. Lett.}, vol.~B52, pp.~344--346, 1974.

\bibitem{Curci:1976yb}
G.~Curci and R.~Ferrari, ``{An Alternative Approach to the Proof of Unitarity
  for Gauge Theories},'' {\em Nuovo Cim.}, vol.~A35, p.~273, 1976.

\bibitem{Ferrari:2004pd}
R.~Ferrari and A.~Quadri, ``{Physical unitarity for massive non-Abelian gauge
  theories in the Landau gauge: Stueckelberg and Higgs},'' {\em JHEP}, vol.~11,
  p.~019, 2004.

\bibitem{tHooft:1972tcz}
G.~'t~Hooft and M.~J.~G. Veltman, ``{Regularization and Renormalization of
  Gauge Fields},'' {\em Nucl. Phys.}, vol.~B44, pp.~189--213, 1972.

\bibitem{Lee:1973fn}
B.~W. Lee and J.~Zinn-Justin, ``{Spontaneously Broken Gauge Symmetries Part 4:
  General Gauge Formulation},'' {\em Phys. Rev.}, vol.~D7, pp.~1049--1056,
  1973.

\bibitem{Grassi:1995wr}
P.~A. Grassi, ``{Stability and renormalization of Yang-Mills theory with
  background field method: A Regularization independent proof},'' {\em Nucl.
  Phys.}, vol.~B462, pp.~524--550, 1996.

\bibitem{Kraus:1997bi}
E.~Kraus, ``{Renormalization of the Electroweak Standard Model to All
  Orders},'' {\em Annals Phys.}, vol.~262, pp.~155--259, 1998.

\bibitem{Glashow:1961tr}
S.~L. Glashow, ``{Partial Symmetries of Weak Interactions},'' {\em Nucl.
  Phys.}, vol.~22, pp.~579--588, 1961.

\bibitem{Weinberg:1967tq}
S.~Weinberg, ``{A Model of Leptons},'' {\em Phys. Rev. Lett.}, vol.~19,
  pp.~1264--1266, 1967.

\bibitem{Salam:1964ry}
A.~Salam and J.~C. Ward, ``{Electromagnetic and weak interactions},'' {\em
  Phys. Lett.}, vol.~13, pp.~168--171, 1964.

\bibitem{Hollik:2002mv}
W.~Hollik, E.~Kraus, M.~Roth, C.~Rupp, K.~Sibold, and D.~Stockinger,
  ``{Renormalization of the minimal supersymmetric standard model},'' {\em
  Nucl. Phys.}, vol.~B639, pp.~3--65, 2002.

\bibitem{Chatrchyan:2012xdj}
S.~Chatrchyan {\em et~al.}, ``{Observation of a new boson at a mass of 125 GeV
  with the CMS experiment at the LHC},'' {\em Phys. Lett.}, vol.~B716,
  pp.~30--61, 2012.

\bibitem{Aad:2012tfa}
G.~Aad {\em et~al.}, ``{Observation of a new particle in the search for the
  Standard Model Higgs boson with the ATLAS detector at the LHC},'' {\em Phys.
  Lett.}, vol.~B716, pp.~1--29, 2012.

\bibitem{Vagnoni:2019muz}
V.~M. Vagnoni, ``{Experimental summary: QCD session of the 54th Rencontres de
  Moriond (Moriond QCD 2019)},'' in {\em {54th Rencontres de Moriond on QCD and
  High Energy Interactions (Moriond QCD 2019) La Thuile, Italy, March 23-30,
  2019}}, 2019.

\bibitem{deFlorian:2016spz}
D.~de~Florian {\em et~al.}, ``{Handbook of LHC Higgs Cross Sections: 4.
  Deciphering the Nature of the Higgs Sector},'' 2016.

\bibitem{Buchmuller:1985jz}
W.~Buchmuller and D.~Wyler, ``{Effective Lagrangian Analysis of New
  Interactions and Flavor Conservation},'' {\em Nucl. Phys.}, vol.~B268,
  pp.~621--653, 1986.

\bibitem{Grzadkowski:2010es}
B.~Grzadkowski, M.~Iskrzynski, M.~Misiak, and J.~Rosiek, ``{Dimension-Six Terms
  in the Standard Model Lagrangian},'' {\em JHEP}, vol.~10, p.~085, 2010.

\bibitem{Jenkins:2013zja}
E.~E. Jenkins, A.~V. Manohar, and M.~Trott, ``{Renormalization Group Evolution
  of the Standard Model Dimension Six Operators I: Formalism and lambda
  Dependence},'' {\em JHEP}, vol.~10, p.~087, 2013.

\bibitem{Jenkins:2013wua}
E.~E. Jenkins, A.~V. Manohar, and M.~Trott, ``{Renormalization Group Evolution
  of the Standard Model Dimension Six Operators II: Yukawa Dependence},'' {\em
  JHEP}, vol.~01, p.~035, 2014.

\bibitem{Alonso:2013hga}
R.~Alonso, E.~E. Jenkins, A.~V. Manohar, and M.~Trott, ``{Renormalization Group
  Evolution of the Standard Model Dimension Six Operators III: Gauge Coupling
  Dependence and Phenomenology},'' {\em JHEP}, vol.~04, p.~159, 2014.

\bibitem{Cheung:2015aba}
C.~Cheung and C.-H. Shen, ``{Nonrenormalization Theorems without
  Supersymmetry},'' {\em Phys. Rev. Lett.}, vol.~115, no.~7, p.~071601, 2015.

\bibitem{Alonso:2014rga}
R.~Alonso, E.~E. Jenkins, and A.~V. Manohar, ``{Holomorphy without
  Supersymmetry in the Standard Model Effective Field Theory},'' {\em Phys.
  Lett.}, vol.~B739, pp.~95--98, 2014.

\bibitem{Elias-Miro:2014eia}
J.~Elias-Miro, J.~R. Espinosa, and A.~Pomarol, ``{One-loop non-renormalization
  results in EFTs},'' {\em Phys. Lett.}, vol.~B747, pp.~272--280, 2015.

\bibitem{Gomis:1995jp}
J.~Gomis and S.~Weinberg, ``{Are nonrenormalizable gauge theories
  renormalizable?},'' {\em Nucl. Phys.}, vol.~B469, pp.~473--487, 1996.

\bibitem{Gomis:1994he}
J.~Gomis, J.~Paris, and S.~Samuel, ``{Antibracket, antifields and gauge theory
  quantization},'' {\em Phys. Rept.}, vol.~259, pp.~1--145, 1995.

\bibitem{Binosi:2019olm}
D.~Binosi and A.~Quadri, ``{Off-shell renormalization in the presence of
  dimension 6 derivative operators. I. General theory},'' 2019.

\bibitem{Binosi:2019nwz}
D.~Binosi and A.~Quadri, ``{Off-shell renormalization in the presence of
  dimension 6 derivative operators. II. UV coefficients},'' 2019.

\bibitem{Binosi:2017ubk}
D.~Binosi and A.~Quadri, ``{Off-shell renormalization in Higgs effective field
  theories},'' {\em JHEP}, vol.~04, p.~050, 2018.

\bibitem{Quadri:2016wwl}
A.~Quadri, ``{Higgs Potential from Derivative Interactions},'' {\em Int. J.
  Mod. Phys.}, vol.~A32, no.~16, p.~1750089, 2017.

\bibitem{Piguet:1995er}
O.~Piguet and S.~P. Sorella, ``{Algebraic renormalization: Perturbative
  renormalization, symmetries and anomalies},'' {\em Lect. Notes Phys.
  Monogr.}, vol.~28, pp.~1--134, 1995.

\bibitem{Ferrari:1999nj}
R.~Ferrari, P.~A. Grassi, and A.~Quadri, ``{Direct algebraic restoration of
  Slavnov-Taylor identities in the Abelian Higgs-Kibble model},'' {\em Phys.
  Lett.}, vol.~B472, pp.~346--356, 2000.

\bibitem{Grassi:1999tp}
P.~A. Grassi, T.~Hurth, and M.~Steinhauser, ``{Practical algebraic
  renormalization},'' {\em Annals Phys.}, vol.~288, pp.~197--248, 2001.

\bibitem{Grassi:2001zz}
P.~A. Grassi, T.~Hurth, and M.~Steinhauser, ``{The Algebraic method},'' {\em
  Nucl. Phys.}, vol.~B610, pp.~215--250, 2001.

\bibitem{Quadri:2003ui}
A.~Quadri, ``{Slavnov-Taylor parameterization for the quantum restoration of
  BRST symmetries in anomaly free gauge theories},'' {\em JHEP}, vol.~04,
  p.~017, 2003.

\bibitem{Quadri:2003pq}
A.~Quadri, ``{Higher order nonsymmetric counterterms in pure Yang-Mills
  theory},'' {\em J. Phys.}, vol.~G30, p.~677, 2004.

\bibitem{Quadri:2005pv}
A.~Quadri, ``{Slavnov-Taylor parameterization of Yang-Mills theory with massive
  fermions in the presence of singlet axial-vector currents},'' {\em JHEP},
  vol.~06, p.~068, 2005.

\bibitem{Bettinelli:2007tq}
D.~Bettinelli, R.~Ferrari, and A.~Quadri, ``{A Massive Yang-Mills Theory based
  on the Nonlinearly Realized Gauge Group},'' {\em Phys. Rev.}, vol.~D77,
  p.~045021, 2008.

\bibitem{Bettinelli:2007cy}
D.~Bettinelli, R.~Ferrari, and A.~Quadri, ``{One-loop self-energy and
  counterterms in a massive Yang-Mills theory based on the nonlinearly realized
  gauge group},'' {\em Phys. Rev.}, vol.~D77, p.~105012, 2008.
\newblock [Erratum: Phys. Rev.D85,129901(2012)].

\bibitem{Bettinelli:2008ey}
D.~Bettinelli, R.~Ferrari, and A.~Quadri, ``{The SU(2) X U(1) Electroweak Model
  based on the Nonlinearly Realized Gauge Group},'' {\em Int. J. Mod. Phys.},
  vol.~A24, pp.~2639--2654, 2009.
\newblock [Erratum: Int. J. Mod. Phys.A27,1292004(2012)].

\bibitem{Bettinelli:2008qn}
D.~Bettinelli, R.~Ferrari, and A.~Quadri, ``{The SU(2) x U(1) Electroweak Model
  based on the Nonlinearly Realized Gauge Group. II. Functional Equations and
  the Weak Power-Counting},'' {\em Acta Phys. Polon.}, vol.~B41, pp.~597--628,
  2010.
\newblock [Erratum: Acta Phys. Polon.B43,483(2012)].

\bibitem{Anselmi:2012qy}
D.~Anselmi, ``{Master Functional And Proper Formalism For Quantum Gauge Field
  Theory},'' {\em Eur. Phys. J.}, vol.~C73, no.~3, p.~2363, 2013.

\bibitem{Anselmi:2012jt}
D.~Anselmi, ``{A Master Functional For Quantum Field Theory},'' {\em Eur. Phys.
  J.}, vol.~C73, no.~4, p.~2385, 2013.

\bibitem{Anselmi:2012aq}
D.~Anselmi, ``{A General Field-Covariant Formulation Of Quantum Field
  Theory},'' {\em Eur. Phys. J.}, vol.~C73, no.~3, p.~2338, 2013.

\bibitem{Ferrari:2005va}
R.~Ferrari and A.~Quadri, ``{A Weak power-counting theorem for the
  renormalization of the non-linear sigma model in four dimensions},'' {\em
  Int. J. Theor. Phys.}, vol.~45, pp.~2497--2515, 2006.

\bibitem{Quadri:2006hr}
A.~Quadri, ``{The Abelian embedding formulation of the Stuckelberg model and
  its power-counting renormalizable extension},'' {\em Phys. Rev.}, vol.~D73,
  p.~065024, 2006.

\bibitem{Quadri:2002nh}
A.~Quadri, ``{Algebraic properties of BRST coupled doublets},'' {\em JHEP},
  vol.~05, p.~051, 2002.

\bibitem{Barnich:2000zw}
G.~Barnich, F.~Brandt, and M.~Henneaux, ``{Local BRST cohomology in gauge
  theories},'' {\em Phys. Rept.}, vol.~338, pp.~439--569, 2000.

\bibitem{Becchi:1985bd}
C.~Becchi, ``{LECTURES ON THE RENORMALIZATION OF GAUGE THEORIES},'' in {\em
  {Relativity, groups and topology: Proceedings, 40th Summer School of
  Theoretical Physics - Session 40: Les Houches, France, June 27 - August 4,
  1983, vol. 2}}, pp.~787--821, 1985.

\bibitem{Wess:1971yu}
J.~Wess and B.~Zumino, ``{Consequences of anomalous Ward identities},'' {\em
  Phys. Lett.}, vol.~37B, pp.~95--97, 1971.

\bibitem{Picariello:2001ri}
M.~Picariello and A.~Quadri, ``{Refined chiral Slavnov-Taylor identities:
  Renormalization and local physics},'' {\em Int. J. Theor. Phys.}, vol.~41,
  pp.~393--408, 2002.

\end{thebibliography}
\bibliographystyle{ieeetr}

\end{document}